\documentclass[11pt]{article}
\usepackage{graphicx}
\usepackage{epsfig}

\newcommand{\BABARPubYear}    {06}

\newcommand{\BABARConfNumber} {005}
\newcommand{\SLACPubNumber} {12017}

\input pubboard/babarsym

\long\def\inst#1{\par\nobreak\kern 4pt\nobreak
    {\it #1}\par\vskip 10pt plus 3pt minus 3pt}

\newcommand{\thv}{\theta_V}
\newcommand{\thl}{\theta_e}
\newcommand{\mkk}{m_{KK}}

\newcommand{\Do}{D^0}

\newcommand{\pipl}{\pi^{+}}

\newcommand{\GeV}{\rm{GeV}}
\newcommand{\GeVc}{\rm{GeV/}c}
\newcommand{\GeVcd}{\rm{GeV/}c^2}
\newcommand{\MeV}{\rm{MeV}}

\newcommand{\ba}{\begin{array}}
\newcommand{\ea}{\end{array}}
\newcommand{\bc}{\begin{center}}
\newcommand{\ec}{\end{center}}
\newcommand{\beq}{\begin{eqnarray}}
\newcommand{\eeq}{\end{eqnarray}}
\newcommand{\bes}{\begin{eqnarray*}}
\newcommand{\ees}{\end{eqnarray*}}
\newcommand{\Zz}{\ifmmode {\rm Z} \else ${\rm Z } $ \fi}
\newcommand{\xxbar}{\ifmmode {\rm x\bar{x}} \else ${\rm x\bar{x}} $ \fi}
\newcommand{\rphi}{\ifmmode {\rm R\phi} \else ${\rm R\phi} $ \fi}

\begin{document}
{\pagestyle{empty}

\begin{flushright}
\babar-CONF-\BABARPubYear/\BABARConfNumber \\
SLAC-PUB-\SLACPubNumber \\
July 2006 \\
\end{flushright}

\par\vskip 5cm

\begin{center}
\Large \bf Measurement of the Hadronic Form Factors in 
\boldmath $\Ds \rightarrow \phi e^+ \nu_e$ Decays 
\end{center}
\bigskip

\begin{center}
\large The \babar\ Collaboration\\
\mbox{ }\\
\today
\end{center}
\bigskip \bigskip

\begin{center}
\large \bf Abstract
\end{center}
Based on the measured four-dimensional rate for $D_{s}^+ \rightarrow \phi e^+ \nu_e$ 
decays, we have determined the ratios of the three
hadronic form factors,
\beq
r_V=V(0)/A_1(0)=1.636 \pm0.067 \pm0.038  ~{\rm and}~r_2=A_2(0)/A_1(0)=0.705 \pm0.056 \pm0.029,
\nonumber
\eeq
using a simple pole ansatz for the $q^2$ dependence, with fixed
values of the pole masses for both the vector and axial form factors. 
By a separate fit to the same data, we have also extracted the pole 
mass for the axial form factors, $m_A$:
\beq
r_V=V(0)/A_1(0)=1.633 \pm0.081 \pm 0.068 ,~r_2=A_2(0)/A_1(0)=0.711 \pm0.111 \pm 0.096\nonumber \\
~{\rm and}~m_A=(2.53^{+0.54}_{-0.35}\pm0.54){\rm{GeV/}c^2}.~~~~~~~~~~~~~~~~~~~~~~~~~~~~~
\nonumber
\eeq

\vfill
\begin{center}

Submitted to the 33$^{\rm rd}$ International Conference on High-Energy Physics, ICHEP 06,\\
26 July---2 August 2006, Moscow, Russia.

\end{center}

\vspace{1.0cm}
\begin{center}
{\em Stanford Linear Accelerator Center, Stanford University, 
Stanford, CA 94309} \\ \vspace{0.1cm}\hrule\vspace{0.1cm}
Work supported in part by Department of Energy contract DE-AC03-76SF00515.
\end{center}

\newpage
} 

\begin{center}
\small

The \babar\ Collaboration,
\bigskip

%
{B.~Aubert,}
{R.~Barate,}
{M.~Bona,}
{D.~Boutigny,}
{F.~Couderc,}
{Y.~Karyotakis,}
{J.~P.~Lees,}
{V.~Poireau,}
{V.~Tisserand,}
{A.~Zghiche}
\inst{Laboratoire de Physique des Particules, IN2P3/CNRS et Universit\'e de Savoie,
 F-74941 Annecy-Le-Vieux, France }
% ...

\end{center}\newpage

\section{INTRODUCTION}
\label{sec:Introduction}

Detailed studies of the dynamics of semileptonic decays
$D \rightarrow V e^+ \nu_e$, where $V$ is a vector meson,
have been performed for non-strange $D$ mesons. 
It is expected that the corresponding semileptonic 
decay of the $D_s$ meson, $\Ds \rightarrow \phi e^+ \nu_e$,~\footnote{Charge 
conjugate states are implied throughout this analysis.}
has similar properties.  So far, 
measurements of this decay have been limited by the size of the 
available data sample.
In this paper, we present a study of the hadronic form factors 
for the decay $\Ds \rightarrow \phi e^+ \nu_e$ with $\phi \to K^+K^-$.

\begin{figure}[htbp]
  \begin{center}
    \mbox{\epsfig{file=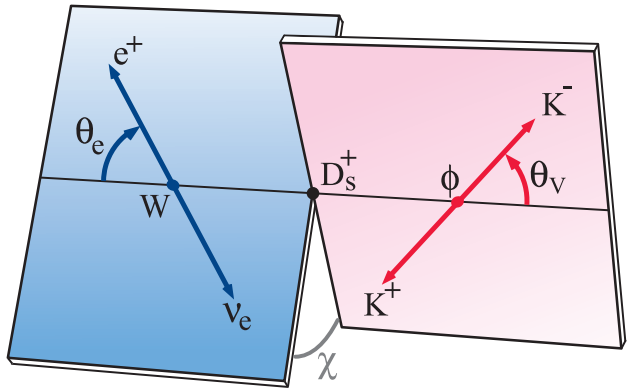,width=0.50\textwidth}}
  \end{center}
  \caption[]{ Definition of the angles $\theta_e$, $\theta_V$, and $\chi$.}
   \label{fig:dsphienu_decay}
\end{figure}

Neglecting the electron mass,
the differential semileptonic decay rate of a scalar meson to a vector meson, specifically, $\Ds \rightarrow \phi e^+ \nu_e$,  
depends on four variables~\cite{ref:KS}
(see Figure \ref{fig:dsphienu_decay}),
\begin{itemize}
\item $q^2$, the invariant mass squared of the $e^+$ and $\nu_e$ ;
\item $\theta_e$, the angle between the direction of the $e^+$ and the
virtual $W^+$, in the $W^+$ rest frame;
\item $\theta_V$, the angle between the direction of the $K^-$ and the
 $\phi$ meson, in the $\phi$ rest frame;
\item $\chi$, the angle between the two decay planes of the $W^+$ and
of the $\phi$, in the $\Ds$ rest frame. It corresponds to the 
angle between the directions of the $e^+$ and of the $K^-$, projected
on a plane normal to the axis defined by the $W^+/\phi$ momentum 
in the $\Ds$ rest frame. $\chi$ is defined in the range from $-\pi$ to $+\pi$.
\end{itemize}
\noindent
It is assumed that $\phi$ decay to $K^+K^-$ is well isolated from decays of other mesons to the same final state, and that any dependence of the rate on the variation of the $K^+K^-$ invariant mass can be neglected. 

The differential decay rate can be written in terms of these variables as follows:
\begin{eqnarray}
{d^4 \Gamma \over d\qsq~d\cos\thv~d\cos\thl~d\chi}
\propto p_{\phi} \qsq \left( \begin{array}{l}
 (1 + \cos \thl )^2\sin^2 \thv |H_ +|^2   \\
  + \,(1 - \cos \thl )^2\sin^2 \thv  |H_ -|^2   \\
  + \,4\sin^2 \thl \cos^2 \thv |H_0|^2  \\
  -\,4\sin \thl (1 + \cos \thl )\sin \thv \cos \thv \cos \chi  H_ + H_0 \\
  +\,4\sin \thl (1 - \cos \thl )\sin \thv \cos \thv \cos \chi  H_ - H_0 \\
  -\,2\sin^2 \thl \sin^2 \thv \cos 2\chi  H_ +  H_ -\\
 \end{array} 
\right)
\label{amp1}
\end{eqnarray}
\noindent 
where $p_{\phi}$ is the momentum of the $\phi$ meson in the rest frame
of the $D_s^+$.  The helicity form factors
can be written in the form
\begin{eqnarray*}
H_\pm (\qsq) &=& (m_D+m_{\phi})A_1(\qsq)\mp 2{m_D~ p_{\phi}\over m_D+m_{\phi}}V(\qsq)\\
H_0 (\qsq) &=& {1\over 2 m_{\phi}\sqrt{\qsq}}
\left[ (m^2_D -m^2_{\phi}-\qsq)(m_D+m_{\phi})A_1(\qsq)
-4{m^2_D~ p_{\phi}^2\over m_D+m_{\phi}}A_2(\qsq) \right]. \\
\end{eqnarray*}

\noindent
$m_D$ and $m_{\phi}$ are the $\Ds$  and $\phi$ masses, respectively.
The vector and axial form factors are generally parameterized 
using an expression based on pole dominance \cite{ref:ks2}:
\begin{equation}
A_i(\qsq)={A_i(0)\over 1-\qsq/m_A^2}~(i=1,2),~~~~~~~
V(\qsq)={V(0)\over 1-\qsq/m_V^2}
\label{eq:pole}
\end{equation}
with the pole masses $m_A = 2.5~\GeVcd$ and $m_V = 2.1~\GeVcd$. 
These values are naive expectations
assuming that the lower mass $c\overline{s}$ states with
$J^P=1^+~{\rm and}~1^-$ dominate the $q^2$ dependence of $A_{1,2}$ and $V$
\footnote{The $1^-$ state contributing to $V$ is the $D_s^*$ of
mass $2.112$ $\GeVcd$,  whereas the $1^+$ states $D_{sJ}(2459)$
and $D_{s1}(2536)$ contribute to $A_{1,2}$}. 
It is expected that the simple pole ansatz has to be modified to include 
contributions from higher mass resonances in addition to the leading 
contribution.
Measurements have usually been expressed in terms of the ratios
of the form factors at $\qsq=0$, namely:
\beq
r_V~=~V(0)/A_1(0)~~{\rm and}~~r_2~=~A_2(0)/A_1(0).
\eeq
At present, there is no experimental determination of $m_A$ and $m_V$.

It has been shown experimentally for the decay 
$\Do \rightarrow K^- e^+ \nu_e$~\cite{ref:bad1401} that the pole ansatz 
(Equation~\ref{eq:pole}) using the nominal values
of the $D_s^*$ mass ($2.112~\GeVcd$) does not provide a good description 
of the $q^2$ dependence of the decay rate.
A fit to data based on this ansatz results in a lower pole mass value, 
$m_V=1.854\pm0.016\pm0.020$ $\GeVcd$.

Becirevic and Kaidalov ~\cite{ref:bk} proposed a modification of the simple pole ansatz for the 
single form factor for semileptonic B and D mesons to pseudoscalar mesons. 
This proposed ansatz has been generalized \cite{ref:svetlana} to describe 
$B$ and $D$ semileptonic decays to vector mesons.  Specifically, 
the three form factors are parameterized as:
\beq
V(\qsq)~=~ \frac{c_H^{\prime}(1-a)}{\left ( 1-\frac{\qsq}{m_{D_s^*}^2} \right)
\left ( 1-a \frac{\qsq}{m_{D_s^*}^2}\right )},
\eeq
\beq
A_1(\qsq)~=~ \xi\frac{c_H^{\prime}(1-a)}{
\left ( 1-b^{\prime}\frac{\qsq}{ m_{D_s^*}^2}\right )},
\eeq
and
\beq
A_2(\qsq)~=~ \frac{c_H^{\prime\prime\prime}}{\left ( 1-b^{\prime}\frac{\qsq}{m_{D_s^*}^2} \right)
\left ( 1-b^{\prime\prime}\frac{\qsq}{ m_{D_s^*}^2}\right )}.
\eeq
Based on this parameterization, $r_V$ is a constant depending only on particle masses,
\beq
r_V~=~\frac{1}{\xi}~=~\frac{\left ( m_{D_s}+m_{\phi}\right )^2}{m_{D_s}^2+m_{\phi}^2}~=~1.8.
\label{eq:svetla}
\eeq

\section{THE \babar\ DETECTOR AND DATASET}
\label{sec:babar}

The data used in this analysis were collected with the \babar\ detector
at the \pep2\ storage rings operating at a center-of-mass (c.m.) energy optimized for
$\FourS$ production.
The \babar\ detector is described in detail elsewhere~\cite{ref:babar}.

This analysis is based on a fraction of the total available \babar\ data sample, 
corresponding to integrated luminosities of $78.5~\fb^{-1}$ recorded on the $\FourS$ resonance.
Samples of Monte Carlo (MC) simulated $\FourS \rightarrow \BB$ decays, the production of charm- and light-quark pairs, equivalent to $4.1,~1.4~{\rm and}~1.1$ times the data statistics have been used
to evaluate the efficiencies and background contributions. A dedicated sample
of simulated signal events, with a uniform phase space distribution and
equivalent to seven times the data, has been used to extract the fitted signal parameters.

\section{ANALYSIS METHOD}
\label{sec:Analysis}

This analysis focuses on semileptonic decays of $\Ds$ mesons which are produced via $e^+ e^- \to c\overline{c}$ annihilation.  $D_s$ mesons produced in $\BB$ events are not included and treated as background.

\subsection{Candidate selection and background rejection}

Fragmentation of the $c$ and the $\overline{c}$ quarks leads to the formation of two jets,
back-to-back in the c.m. frame. In most cases, each jet contains one charm meson. 
The event thrust axis is determined from all charged and neutral particles
measured in the c.m. system.  To minimize the loss of particles close to the beam axis and to ensure a good reconstruction of the total energy and momentum in the event, we select events for which the direction of the thrust 
axis is in the interval $|\cos(\theta_{thrust})|<0.75$.

Three variables, $R_2$ (the ratio of the second and zeroth
order Fox-Wolfram moments \cite{ref:r2}), the total 
multiplicity of charged and neutral particles,
and the momentum of the fastest track in the event
are used to reduce the contribution from $\BB$ events.
These variables have been combined linearly to form a 
Fisher discriminant, ${\cal F}$. We choose a cut on
${\cal F}$ that retains $68\%$ of signal events and removes 70$\%$ of 
$\BB$ background.

A plane perpendicular to the thrust axis is used to define two hemispheres, 
equivalent to the two jets produced by quark fragmentation.
In each hemisphere,  we search for decay products of the $\Ds$, a charged lepton and two oppositely charged kaons. We use as charged leptons only positrons (or 
electrons for the charge conjugate $D_s^-$ decays)
with a c.m. momentum larger than 0.5 $\GeVc$.

Since the neutrino ($\nu_e$) momentum is unmeasured, a kinematic
fit is performed, constraining the invariant mass of the candidate $(e^+ \Kp \Km \nu_e)$ system to the $\Ds$ mass. In this fit, the $\Ds$ momentum and the neutrino energy 
are estimated from the other particles measured in the event.
The $\Ds$ direction is taken as the direction opposite to the sum
of the momenta of all reconstructed particles in the event, except for the 
two kaons and the positron associated with the signal candidate. 
The neutrino energy is estimated as the difference between the total energy of the jet
and sum of the energies of all reconstructed particles in that hemisphere.
The energy of the jet is determined from its mass and momentum.  The jet mass is constrained taking into account that each jet 
contains at least one charm particle and thus its mass has to exceed 
the charm particle mass.   
The $\Ds$ candidates are retained if the $\chi^2$ probability of the kinematic fit exceeds $10^{-3}$.

Tracks present in the signal hemisphere, which are not decay products
of the $\Ds$ candidate, are referred to as ``spectator'' ($spec$) tracks. 
Since the charm hadrons in a $c$ or $\overline{c}$ jet carry a large
fraction of the jet energy, their decay products
have on average higher energies than spectator particles. 
The following variables are used to define a second Fisher discriminant designed to select the signal $\Ds \rightarrow \phi e^+ \nu_e$ decays,
\begin{itemize}
\item the fitted $\Ds$ momentum ($P_{D_s}$);
\item the mass of the spectator system ($m_{spec.}$);
\item the direction of the spectator momentum relative to the thrust axis ($\cos{(spec.,thrust)}$);
\item the momentum of the leading spectator track ($P_{leading}$), i.e. the 
the spectator track having the largest momentum;
\item the total momentum of the spectator system ($P_{spec}$).
\end{itemize}

\begin{figure}[htbp]
  \begin{center}
 \mbox{
   \epsfig{file=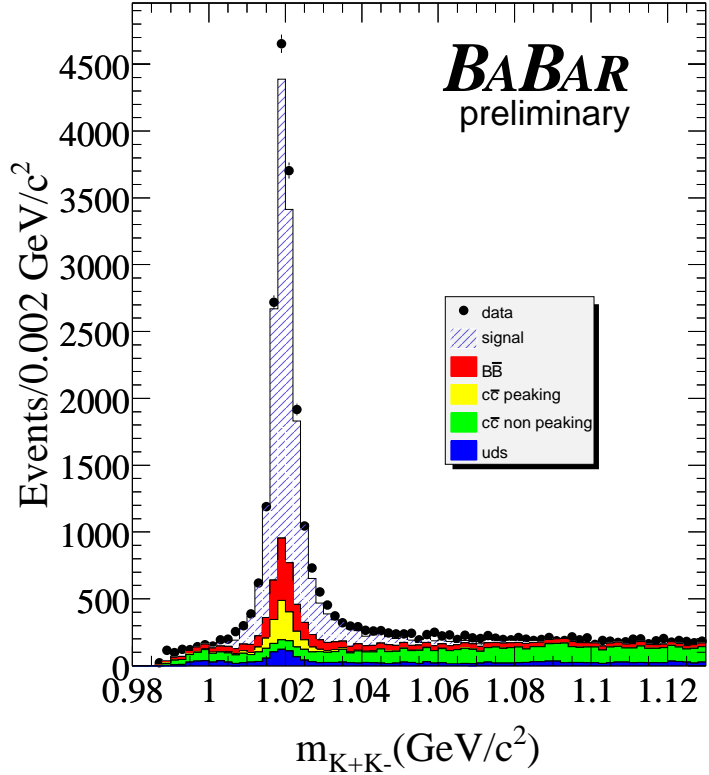,
    width=0.60\textwidth}
  }
  \end{center}
  \caption[]{$K^+K^-$ invariant mass distribution from data and simulated events. MC events have been normalized to the data luminosity according to the different cross sections. The excess of signal events in the $\phi$ region
can be attributed to a different production rate and decay branching fraction
of $\Ds$ mesons in data and in simulated events. Dedicated studies have been
done to evaluate the amount of peaking background in real events.}
   \label{fig:bkg_comp}
\end{figure}

Figure~\ref{fig:bkg_comp} shows the $K^+ K^-$ invariant mass distribution for the selected decays compared to MC simulation and the composition of the background. 
We define $\phi$ candidates as $K^+ K^-$ pairs with an invariant mass 
in the interval from 1.01 and 1.03 $\GeVcd$.
We use a cut on the second Fisher discriminant that retains 64$\%$ of signal events
and rejects 78$\%$ of combinatorial background.
Of the background contribution of 26$\%$ in the signal region, 
$14~\%$ are from continuum $q\overline{q}$ events (with $q=u,d,s$), $23.4~\%$ are  from $\B^0\bar{B}^0$ events, $21.6~\%$ from 
$\Bp\Bm$ events, and the remainder are $c\overline{c}$ events. 
About 71$\%$ of the total background include a true $\phi$ decay combined 
with an electron from another source, namely $B$ meson decays (41$\%$), charm particle decays (25$\%$), photon conversions or Dalitz decays (24$\%$), and the rest are fake electrons.
These $\phi$ mesons are expected to originate from the 
primary vertex, or from a secondary charm decay vertex.

\subsection{Measurement of decay distributions}

Taking into account the results of the kinematic fit, the decay rates are studied as a function of the following variables: $q^2$ ($q^2_r$), $\cos(\thl)$ ($\cos(\thl)_r$), $\cos(\thv)$ ($\cos(\thv)_r$) and 
$\chi$ ($\chi_r$). 
Using simulated events, the resolution for these reconstructed variables
has been  
studied by comparing the reconstructed (indicated by the index $r$) and true values.

The resolution functions have been fitted by the sum of two Gaussian distributions. The fitted standard deviations are listed in Table~\ref{tab:res}. 
This information is presented here to illustrate the performances of the reconstruction and the kinematic fit.  The resolution parameters are not used in the fit to the decay distributions. Instead, the MC simulation uses the identical reconstruction of the four kinematic variables as used for the data, and thus the distributions of the simulated decays are expected to reproduce the data.

We have chosen a narrow interval in the $K^+K^-$ invariant mass to select $\phi$ meson candidates. Any decay-rate variation as a function of the two-kaon mass is ignored.

\begin{table}[htbp]
 \begin{center}
  \caption[]{Resolution of the reconstructed four kinematic variables:  Standard deviations of the two Gaussian distributions, and their relative contribution.
 \label{tab:res}}

  \begin{tabular}{|c|c|c|c|}	
    \hline	
variable  & $\sigma_1$ & $\sigma_2$ &fraction of the \\
     & & & narrower Gaussian\\
\hline
$q^2$  & $0.0778$ $\GeV^2$ & $0.249$ $\GeV^2$ &0.33 \\
$\cos(\thl)$ & $0.046$ & $0.228$ &0.47 \\
$\cos(\thv)$ & $0.099$ & $0.387$ &0.43 \\
$\chi$ & $0.262$ rad & $1.39$ rad &0.41 \\
    \hline
   \end{tabular}
  \end{center}
\end{table}

\subsection{Fitting procedure}
\label{sec:fits}

We perform  
a maximum likelihood fit to the four-dimensional decay distribution in the 
variables $q^2_r$, 
$\cos(\theta_V)_r$, $\cos(\theta_e)_r$ and $\chi_r$
using the likelihood function
\beq
{\cal L} = - \sum_{i=1}^{\rm nbins} \ln {{\cal P}(n^{\rm data}_i |n^{\rm MC}_i) }.
\eeq
\noindent
In this expression, for each bin $i$, ${\cal P}(n^{\rm data}_i |n^{\rm MC}_i)$ is the
Poisson probability to observe $n^{\rm data}_i$ events, when
$n^{\rm MC}_i$ are expected.
Considering the typical resolutions given
in Table \ref{tab:res} and the available statistics, we have chosen
five bins for each of the four variables, corresponding a four-dimensional array with a total of $\rm nbins=625$.
\\The expected number of events results from:
\begin{itemize}
\item combinatorial background in  the $\phi$ signal interval;

\item peaking background, i.e. real $\phi $ decays combined with a background electron; 

\item $\phi e^+ \nu_e$ signal events.

\end{itemize}

The number of expected signal events is obtained from MC simulation in the following way.  A dedicated sample of signal events is generated with a uniform decay phase space distribution, and each event
is weighted using the differential
decay rate given in Equation~\ref{amp1}, divided by $p_{\phi}$.

We take advantage of the fact that the estimated background rate is flat in two of the four variables, $\cos(\theta_V)$ and $\chi$
(see Figure \ref{fig:fitdata}), by averaging over these distributions,
\beq
n^{\rm bckg.}_{i_{q^2},i_{\rm cos(\thl)},i_{\rm cos(\thv)},i_{\chi}} = 
\frac{\sum_{j,k=1}^{\rm nbin_{\rm cos(\thv)},\rm nbin_{\chi}} 
n^{\rm bckg.}_{i_{q^2},i_{\rm cos(\thl)},j,k}}{\rm nbin_{\rm cos(\thv)}\rm nbin_{\chi} }
\eeq
\noindent
This expression applies to each component of background. 
The background components are normalized to correspond to the 
expected rates for the integrated luminosity of the data sample.

The absolute normalization for signal events $(N_S)$ is left free to vary
in the fit. In each bin $(i)$, the expected number of events is evaluated 
to be:
\beq
n^{\rm MC}_i = N_S \frac{\sum_{j=1}^{n_i^{\rm signal}} w_j(\lambda_k)}{W_{\rm tot}(\lambda_k)}~+~ n^{\rm bckg.}_i.
\eeq
Here $n_i^{\rm signal}$ refers to the number of simulated signal events, with reconstructed values of the four variables corresponding to bin $i$. The weight $w_j$ is evaluated for each event, using the generated values of the 
kinematic variables, thus accounting for resolution effects.
$W_{\rm tot}(\lambda_k)=\sum_{j=1}^{N^{\rm signal}} w_j(\lambda_k)$
is the sum of the weights for all simulated signal events which have been 
generated according to a uniform phase space distribution. $N_S$ and 
$\lambda_k$ are the parameters to be fitted. Specifically, the free parameters   $\lambda_k$ are $r_V$, $r_2$, and parameters which define
$q^2$ dependence of the form factors. To avoid having to introduce finite
ranges for the fit to the pole masses, $m_i$, we define $m_i = 1 +\lambda_i^2$. This expression ensures that $m_i$ is always larger than $q^2_{max.}\simeq 0.9$ $\GeV^2$.

\section{RESULTS OF THE FIT TO THE DECAY RATE}

\begin{figure}[htbp]
  \begin{center}
    \mbox{
   \epsfig{file=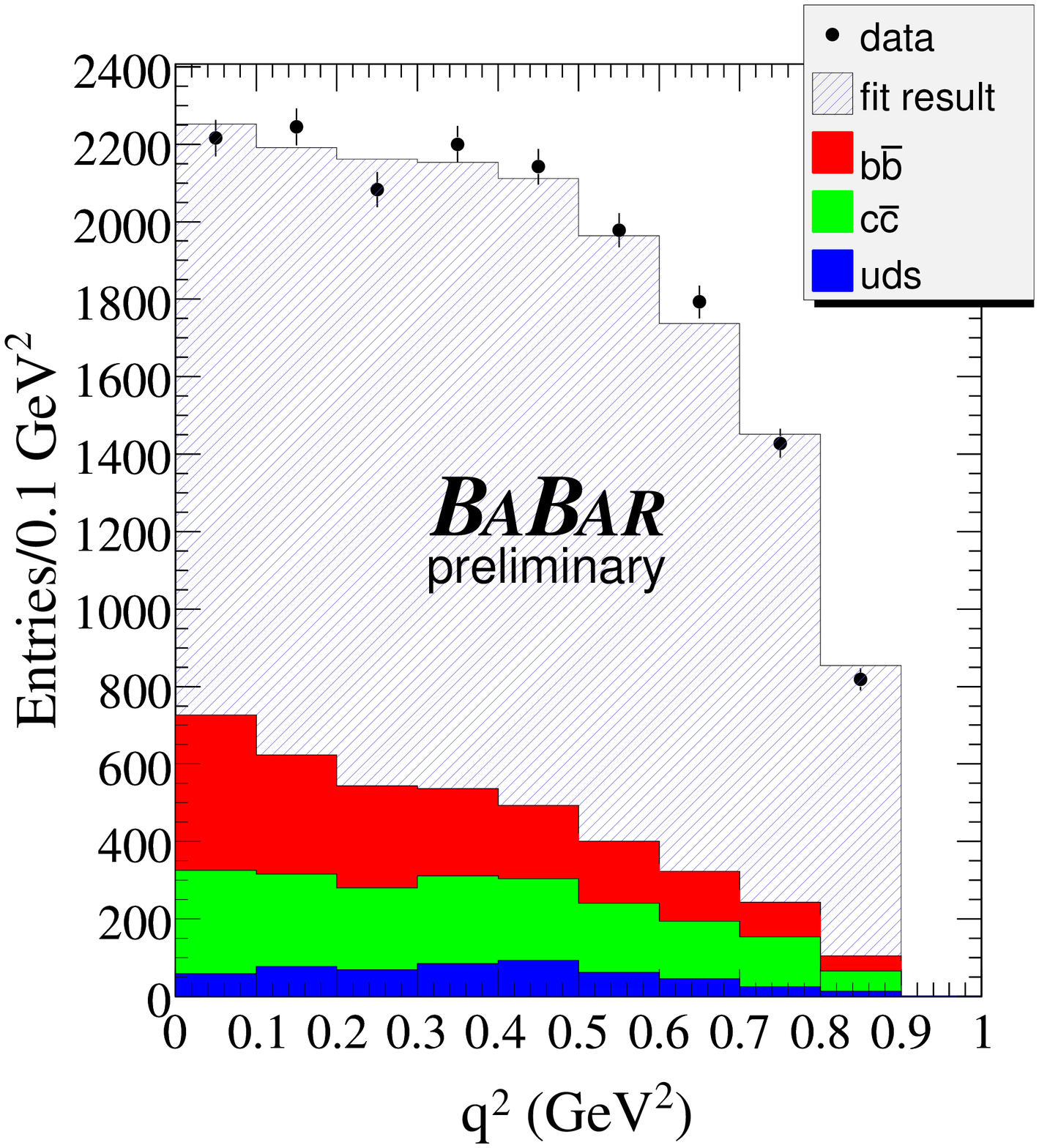,
    width=0.3\textwidth}
   \epsfig{file=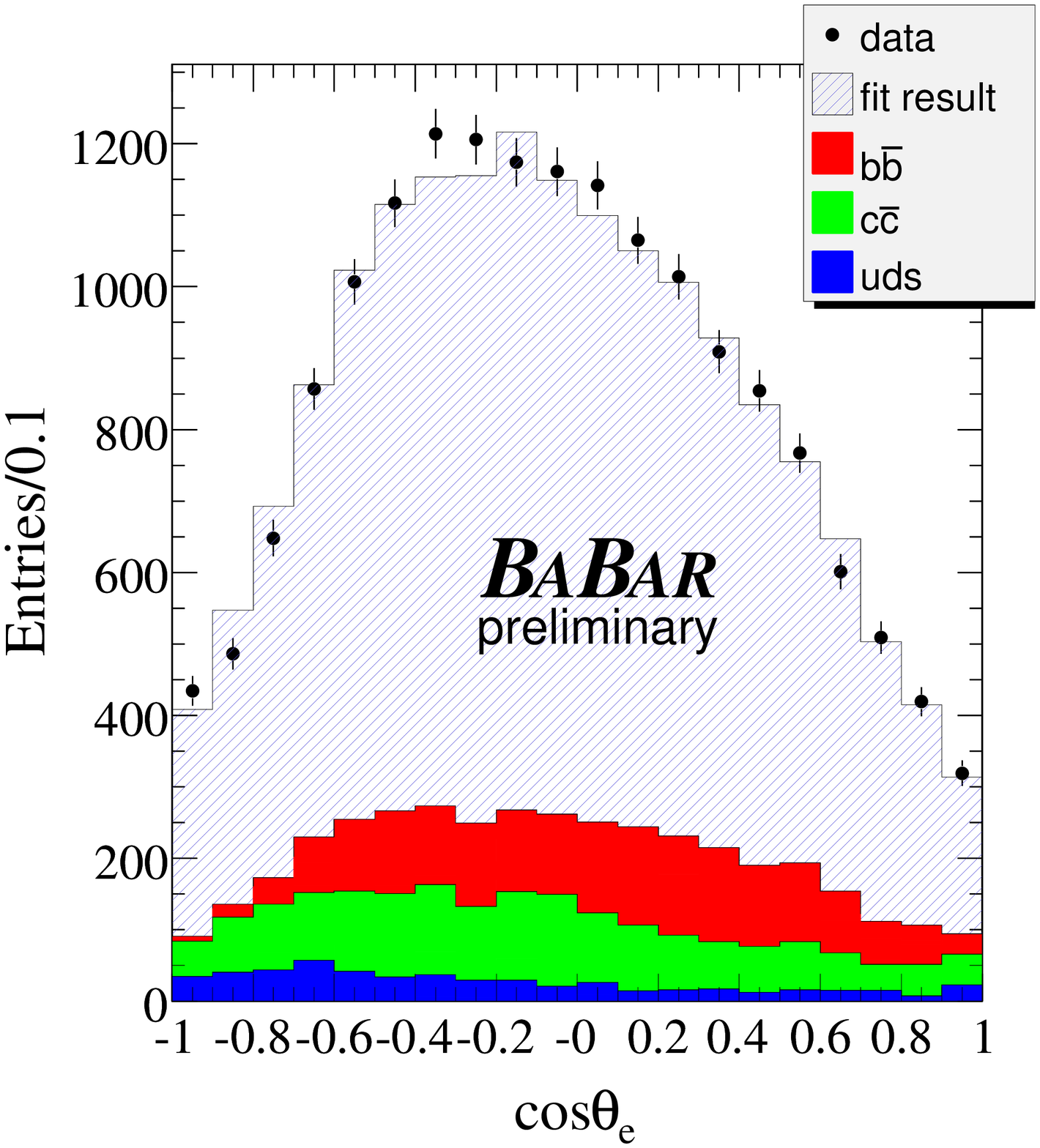,
   width=0.3\textwidth} }

 \mbox{
   \epsfig{file=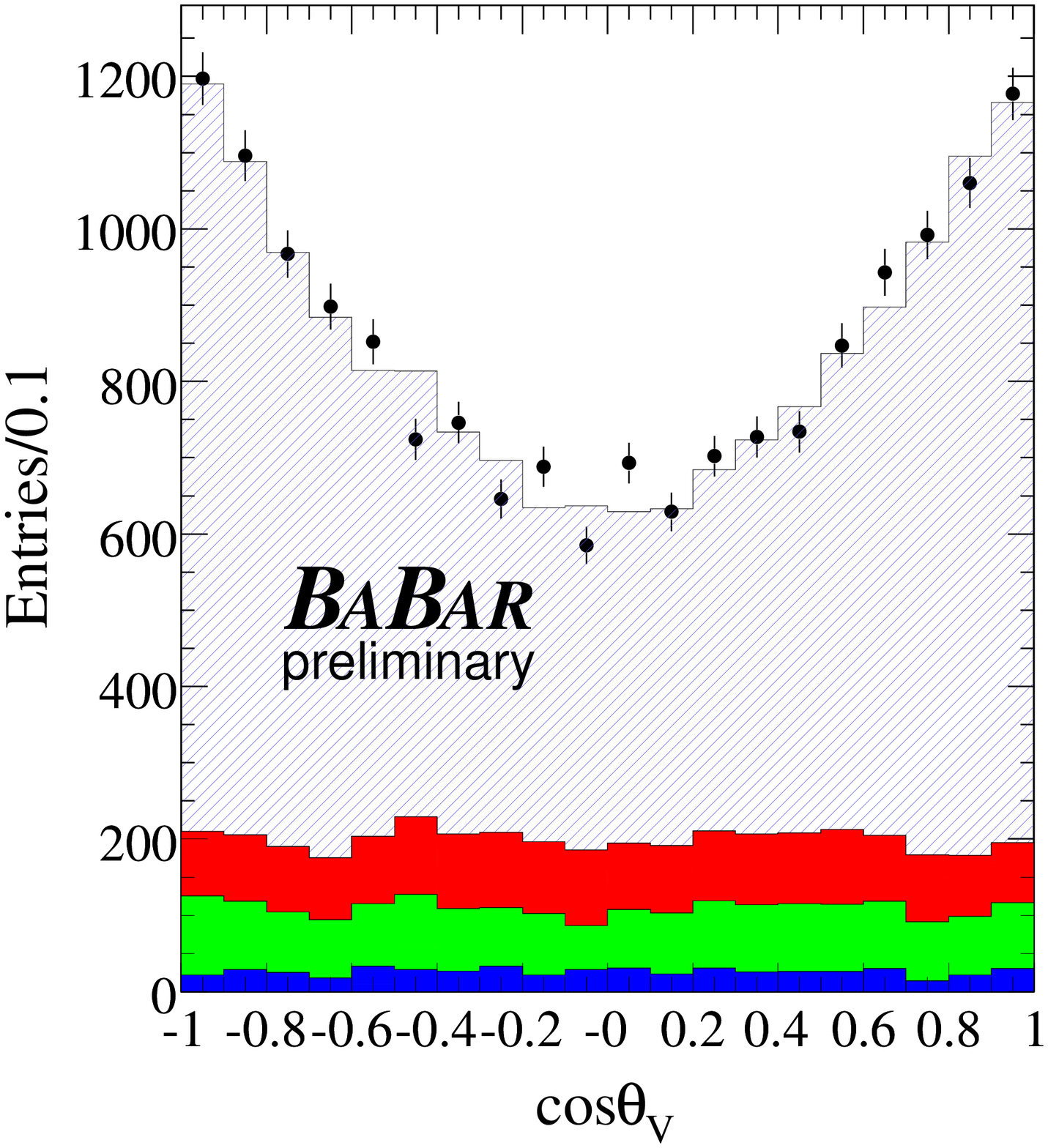,
    width=0.3\textwidth}
   \epsfig{file=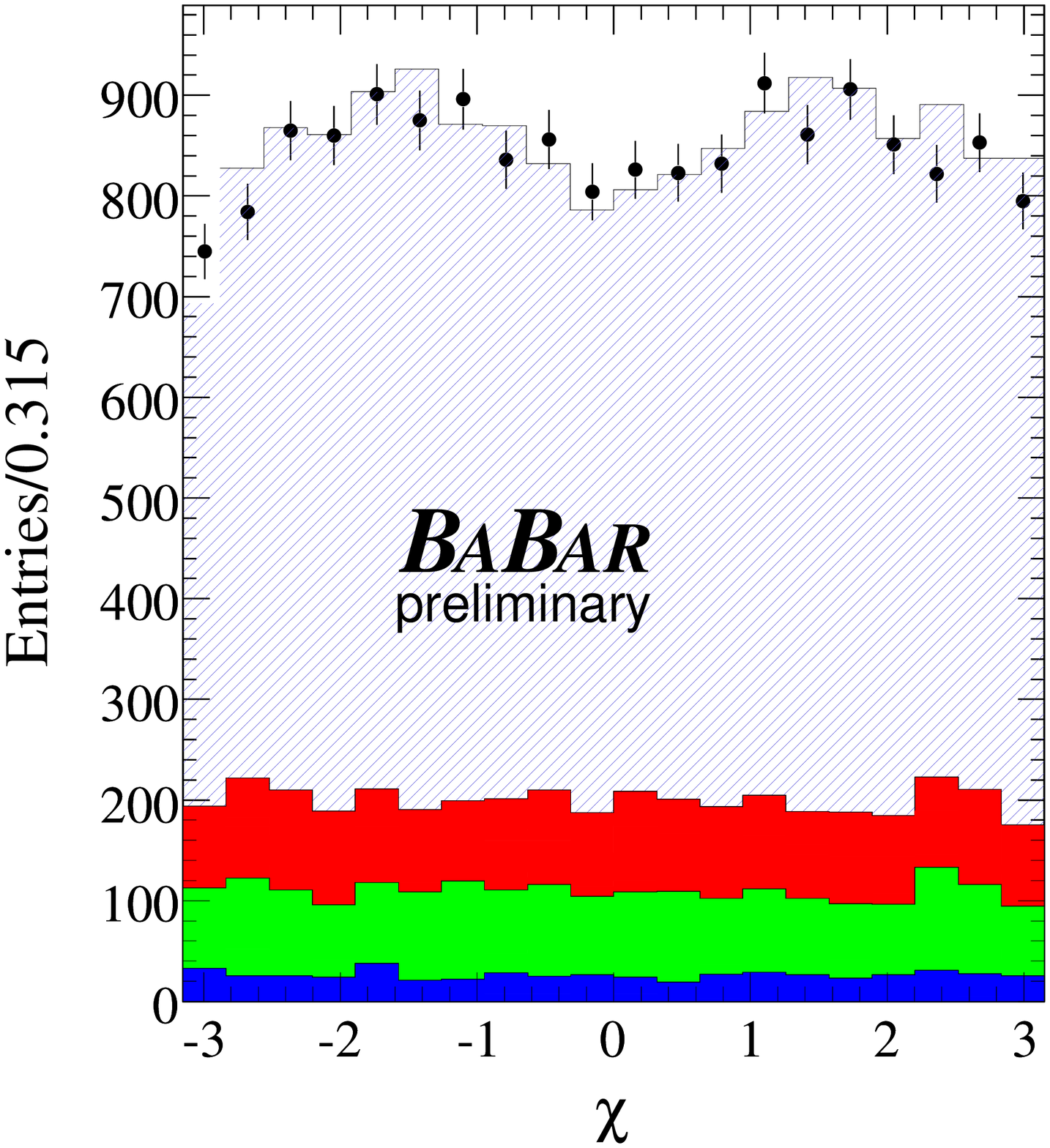,
    width=0.3\textwidth}
  }

  \end{center}
  \caption[]{Projected distributions of the reconstructed four kinematic variables which define the decay rate for
$\Ds \rightarrow \phi e^+ \nu_e$. The data (with statistical errors) are compared to histograms showing size of the fitted signal and background contributions.
}
   \label{fig:fitdata}
\end{figure}

The fit to the four-dimensional data distribution is performed using simulated signal events generated according to a uniform phase space distribution.
Signal MC events are weighted to correct for differences 
in the quark fragmentation process between data and simulated events. 
Using fixed values for the pole masses ($m_A =
2.5~\GeVcd$ and $m_V = 2.1~\GeVcd$) the following results
are obtained:
\beq
N_S=12886 \pm 129,~r_V=1.636\pm0.067,~r_2=0.705\pm 0.056.\nonumber
\eeq
\noindent
The measured distributions, projected on the four variables, 
are compared with the results of the fit in Figure \ref{fig:fitdata}. 
The fit procedure has been verified with fits to a large number of simulated 
toy experiments with event samples of comparable size.

If we keep $m_V$ fixed, the pole mass for the vector form factor, for which there is no sensitivity, and leave $m_A$, the pole mass of the axial vector form factor, as a free parameter, we obtain 
\beq
N_S=12887 \pm 129,~r_V=1.633\pm0.081,~r_2=0.711\pm 0.111,
~m_A=(2.53^{+0.54}_{-0.35})~\GeVcd.\nonumber
\eeq

\section{SYSTEMATIC STUDIES}
The following sources of systematic uncertainties have been considered.

\subsection{Generator tuning}

The fraction of the beam energy carried by a $\Ds$ meson is rather different in data and MC simulation. This difference has been measured 
using $\Ds \rightarrow \phi \pipl$ decays. After applying this correction,
there remain small differences in the distributions of the variables entering
the Fisher discriminant, used to reduce the background
level. These differences have been evaluated  using also samples of $\Ds \rightarrow \phi \pipl$ decays.
The largest of the remaining differences results in the following changes of the fitted parameters,
\beq
\delta(N_S)= +5,~\delta(r_V)=-0.005,~\delta(r_2)=+0.008.
\eeq
For the second fit with variable pole mass, $m_A$, these changes are:
\beq
\delta(N_S)= +3,~\delta(r_V)=+0.008,~\delta(r_2)=-0.017,
~\delta(m_A)=-0.10~\GeVcd.
\eeq

\subsection{Background control}

Two contributions have to be considered:

\subsubsection{Combinatorial background}
The level of combinatorial background has been evaluated using the mass intervals $1.10 < \mkk <1.15~\GeVcd$, where the contributions from true $\phi \to K^+K^-$ decays are negligible.
In this region we found an excess of $7\%$ in data over MC simulated events.  We assign 
a $10\%$ uncertainty to the combinatorial background
estimate. The corresponding systematic uncertainties
on $r_V$ and $r_2$ are:
\beq
\delta(r_V)=\pm0.008,~\delta(r_2)=\pm0.003. \nonumber
\eeq
For the second fit with variable pole mass, $m_A$,  these uncertainties are:
\beq
\delta(r_V)=\pm0.022,~\delta(r_2)=\pm0.032,~\delta(m_A)=\pm0.114~\GeVcd. \nonumber
\eeq

\subsubsection {Background from $\phi$ mesons produced in $D$ or $B$ decays}
The rate of these background sources depends on branching fractions to $\phi$ mesons.
A study has been performed to compare the $\phi$ meson production rate
in data and simulated events using different event samples which
have been normalized to the same integrated luminosity. 
Events with a candidate electron and a candidate $\phi$ are used since they 
correspond to a sample enriched in charm decays.
The production of a $\phi$ meson is studied, in the same hemisphere that contains the lepton, and in the other.

The production of $\phi$ mesons in remaining $\BB$ background has been measured by subtracting off-peak from on-peak events 
\footnote{On-peak events are recorded at the $\FourS$ energy whereas
off-peak events are recorded at an energy 40 $\MeV$ below.}.
From comparisons of data and MC simulated $\BB$ samples, we conclude that the simulation provides
a description of $\phi$ production, accompanying a lepton, with
an accuracy better than 10$\%$.

The production of $\phi$ mesons accompanied by an electron has been also studied using off-peak events. These events
have contributions from $c$ and $u,d,s$ quark-pair production.
Correction factors, to be applied to the simulated $\phi$
rate, are determined such that the total expected $\phi$ production
and the fractions expected from $c$- and
$u,d,s$ events agree with data. The values are given in 
Table \ref{tab:corr_phi}. We assign a systematic uncertainty of $\pm 10\%$
to these corrections.

\begin{table}[htbp]
\begin{center}
\caption[]{Correction factors to be applied on the simulated $\phi$
production rate in $e^+e^- \to q \overline{q}$. 
  \label{tab:corr_phi}}

  \begin{tabular}{|c|c|c|}
    \hline
quark  & $\phi$ accompanying & $\phi$ opposite  \\
         & the lepton          & the lepton \\
\hline
$c$  & $0.85$ & $0.80$  \\
\hline
$uds$  & $1.06$ & $1.05$  \\
    \hline
  \end{tabular}  
\end{center}
\end{table}

Background events from $\Ds \rightarrow \phi \pi^0 e^+ \nu_e$ decays can have decay
characteristics that differ slightly from the signal decays. The $\phi$ and the positron originate from the same $\Ds$ hadron, contrary to other
peaking background sources. But the rate for this decay is suppressed by the OZI rule, and the detection
efficiency is expected to be lower than for the signal events. In the following, its contribution is neglected. 

Considering $\pm$10\% uncertainties on the $B$ and $D$ peaking background,
the corresponding systematic 
uncertainties on $r_V$ and $r_2$ are:
\beq
\delta(r_V)=\pm0.019,~\delta(r_2)=\pm0.009. \nonumber
\eeq
\\If, in addition, $m_A$ has been fitted, the uncertainties are:
\beq
\delta(r_V)=\pm0.052,~\delta(r_2)=\pm0.072,~\delta(m_A)=\pm0.48~\GeVcd. \nonumber
\eeq

\subsection{Monte Carlo statistics}

In the fitting procedure two sources of statistical fluctuations are 
not included. They originate from
the finite statistics of the weighting procedure applied
to simulated signal events and from uncertainties in the estimate of 
the average number of background events in each bin.  
As a result, statistical uncertainties obtained from the standard fits
to data or simulated events may be underestimated and the values
of the fitted parameters may have biases.

The effect of these statistical errors on the fitted parameters $r_V$ 
and $r_2$ have been evaluated using 1000 simulated toy fits. For each toy fit, not only the ``fake experiment'' is generated, but also the sample of pure signal and the background distributions are created. 
In each of these experiments, the same  number of signal events (13 000) and the same ratio background over signal, B/S=0.31 as the data is used.  
The predicted distributions in the fit are generated uniformly over the decay phase space, with samples of 110,000 events each, as in the real fit. 
The  width of the normalized pull distributions of the toy fits is 1.07 and the bias is 0.08. Uncertainties on these numbers are $\pm 0.03$.
The systematic uncertainty attached to possible biases is assumed to be $0.1 \times \sigma_{\rm fit}$, where $\sigma_{\rm {fit}}$ is the
statistical uncertainty of the fit.
The uncertainty assigned to the size of the simulated events sample and to the statistical uncertainties on the average number of
background events in each bin, is estimated from the increase
of the width of the pull distribution relative to unity:
\beq
   \sqrt{1.1^2 -1} \times \sigma_{\rm fit} \simeq 0.46 \times \sigma_{\rm fit},
\label{eq:toy}
\eeq    
\\where 0.1 is chosen as the upper limit for the observed deviations.

\subsection{Remaining detector effects}

Effects induced by momentum dependent differences on the electron
and charged kaon reconstruction
efficiency between data and simulated events have been evaluated.
Standard correction factors determined from selected control data samples, have been applied to correct for these differences which are typically of a few percent. 
The impact of this correction on the fitted parameters are:
\beq
\delta(r_V)=+0.018,~\delta(r_2)=+0.012. \nonumber
\eeq
\\If, in addition, $m_A$ is fitted, the variations are:
\beq
\delta(r_V)=+0.017,~\delta(r_2)=+0.015,~\delta(m_A)=+0.02~\GeVcd. \nonumber
\eeq

The systematic uncertainty of these corrections is estimated to be 30\% and corresponding values have been given in Tables
 \ref{tab:systematics} and \ref{tab:systematics_ma} respectively.

\subsection{Reconstruction accuracy on the kinematic variables}

Using $\Dstarp \rightarrow \Do \pi^+$ and $~\Do  \rightarrow K^-\pi^+\pi^0$
events it has been verified that differences between data and
simulated events in the resolution of the variables $\qsq$
and $\cos(\thl)$ are small compared with other
sources of systematic uncertainties. They have been neglected at present.

\subsection{Summary of systematic uncertainties}
A summary of the systematic uncertainties on the measurement of $r_V$ and $r_2$ is given in Table \ref{tab:systematics} and 
in  Table \ref{tab:systematics_ma}, for fits that include  $m_A$ as a free parameter. 

\begin{table}[htbp]
\begin{center}
 \caption[]{Systematic uncertainties on $r_V$ and $r_2$.
  \label{tab:systematics}}
  \begin{tabular}{|c|c|c|}
    \hline
Source  & error on $r_V$ &   error on $r_2$ \\
\hline
Generator tuning  & $0.005$ & $0.008$  \\
Background control  & $0.021$ & $0.009$  \\
Monte-Carlo statistics  & $0.031$ & $0.026$  \\
Detector effects  & $0.006$ & $0.004$  \\
    \hline
Total & $0.038$ & $0.029$  \\
\hline
  \end{tabular}
\end{center}
\end{table}

\begin{table}[htbp]
\begin{center}
 \caption[]{Systematic uncertainties on $r_V$, $r_2$ and $m_A$.
  \label{tab:systematics_ma}}
  \begin{tabular}{|c|c|c|c|}
    \hline
Source  & error on $r_V$ &   error on $r_2$ &   error on $m_A$\\
   & & & $(\GeVcd)$\\
\hline
Generator tuning  & $0.008$ & $0.017$ & $0.10$ \\
Background control  & $0.056$ & $0.079$ & $0.49$ \\
Monte-Carlo statistics  & $0.038$ & $0.052$ & $0.21$ \\
Detector effects  & $0.006$ & $0.005$ & $0.01$ \\
    \hline
Total & $0.068$ & $0.096$ & $0.54$  \\
\hline
  \end{tabular}
\end{center}
\end{table}

\section{RESULTS AND CONCLUSIONS}

Assuming pole dominance for the different form factors and using the
fixed pole mass values, the contributions of the $A_2$ and $V$ 
hadronic form factors,
relative to $A_1$, have been measured in a sample of 13,000 
$\Ds \rightarrow \phi e^+ \nu_e$ decays. In this measurement,
pole mass expressions have been used for the $q^2$ dependence of the form 
factors and values for the pole masses, equal to those assumed in 
previous experiments have been used. We have obtained:
{
\beq
r_V=V(0)/A_1(0)=1.636 \pm0.067 \pm0.038  ~{\rm and}~r_2=A_2(0)/A_1(0)=0.705 \pm0.056 \pm0.029 \nonumber
\eeq}
\\where the first uncertainty is statistical, and the second is systematic.
These values are compatible with and more accurate than previous determinations.

The present
measurement has a limited sensitivity on $m_V$ and its value has been fixed
at $2.1 ~\GeVcd$. Allowing the pole mass of the axial form factors, $m_A$,
to vary as a free parameter in the fit, we obtain  
{
\beq
r_V=V(0)/A_1(0)=1.633 \pm0.081 \pm 0.068 ,~r_2=A_2(0)/A_1(0)=0.711 \pm0.111 \pm 0.096 \nonumber \\
~{\rm and}~m_A=(2.53^{+0.54}_{-0.35}\pm0.54)~\GeVcd.~~~~~~~~~~~~~~~~~~~~~~~~~~~~~\nonumber
\eeq
\noindent
The fitted value of $m_A$ agrees with the assumed default value.

In Table \ref{tab:meast}, the results of this analysis are compared with earlier measurements obtained in photoproduction experiments
at Fermilab~\cite{ref:e653phi,ref:e687phi,ref:e791phi,ref:focusphi} and
by the CLEOII experiment~\cite{ref:cleophi}. The central values and corresponding total errors are also shown in Figure \ref{fig:r2rv}.

\begin{table}[htbp]
\begin{center}
 \caption[]{Results from previous experiments and present measurements.
They have been obtained assuming a pole mass dependence for the hadronic form
 factors with fixed values of the pole masses: $m_A=2.5~\GeVcd$ and 
$m_V=2.1~\GeVcd$.
  \label{tab:meast}}
  \begin{tabular}{|c|c|c|c|}
    \hline
Experiment & Statistics (S/B)& $r_V$ & $r_2$  \\
\hline
 E653 \cite{ref:e653phi} & 19/5 &$2.3^{+1.1}_{-0.9}\pm0.4$ & $2.1^{+0.6}_{-0.5}\pm0.2$  \\
\hline
 E687 \cite{ref:e687phi} & 90/33 &$1.8 \pm0.9 \pm0.2$ & $1.1 \pm0.8 \pm0.1$  \\
\hline
 CLEOII \cite{ref:cleophi} &308/166  &$0.9 \pm0.6 \pm0.3$ & $1.4 \pm0.5 \pm0.3$  \\
\hline
 E791 \cite{ref:e791phi} & $\sim$300/60 &$2.27 \pm0.35\pm0.22$ & $1.57 \pm0.25\pm0.19$  \\
\hline
 FOCUS \cite{ref:focusphi} &$\sim$560/250  &$1.549 \pm 0.250\pm0.145$ & $0.713 \pm0.202\pm0.266$  \\
\hline \hline
 $\babar$ & 12972/3931 &$1.636\pm0.067\pm0.038$ & $0.705\pm0.056\pm0.029$ \\ 
\hline
  \end{tabular}
\end{center}
\end{table}
 
\begin{figure}[htbp]
  \begin{center}
 \mbox{
   \epsfig{file=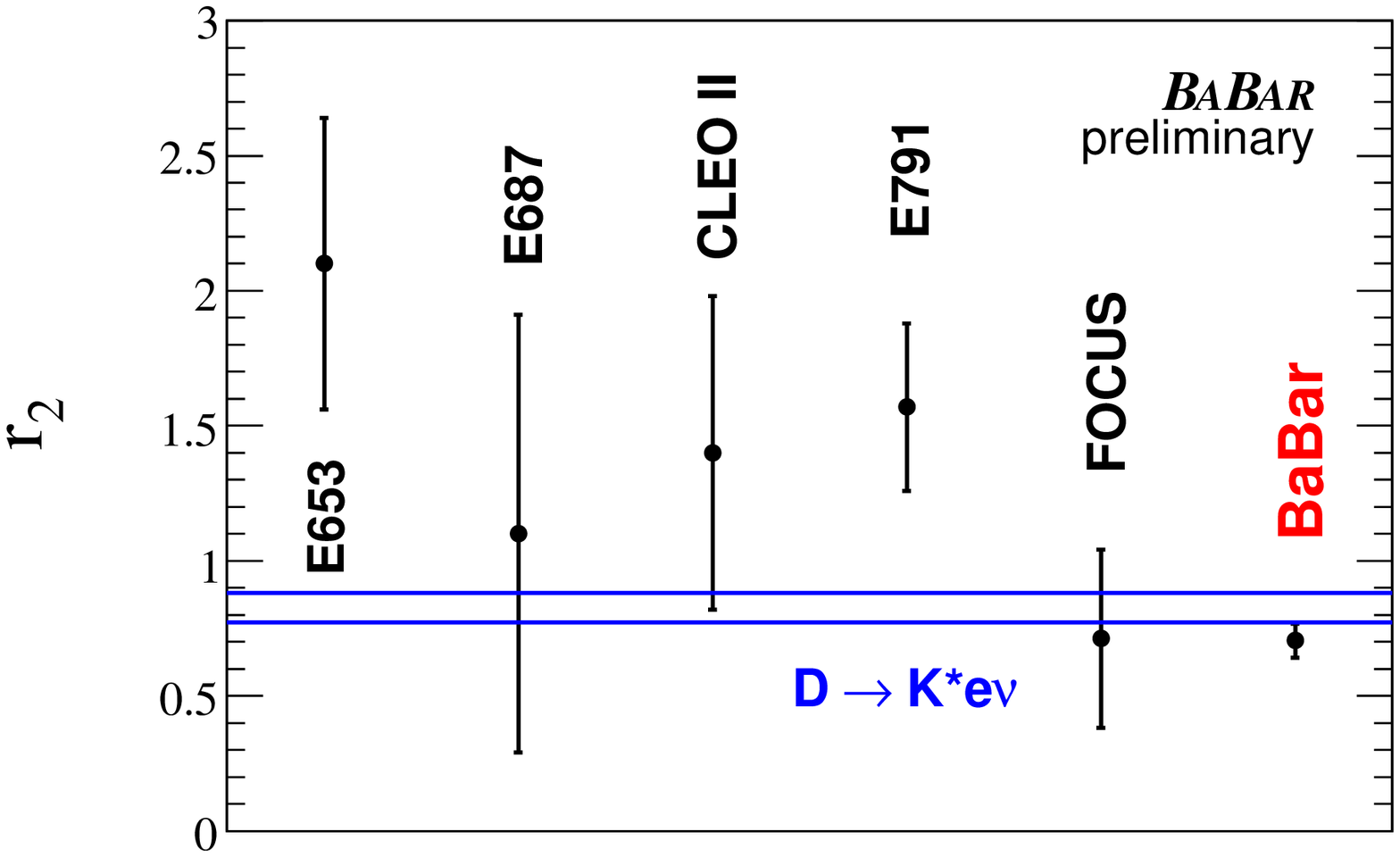,
    width=0.6\textwidth}
  }
 \mbox{
   \epsfig{file=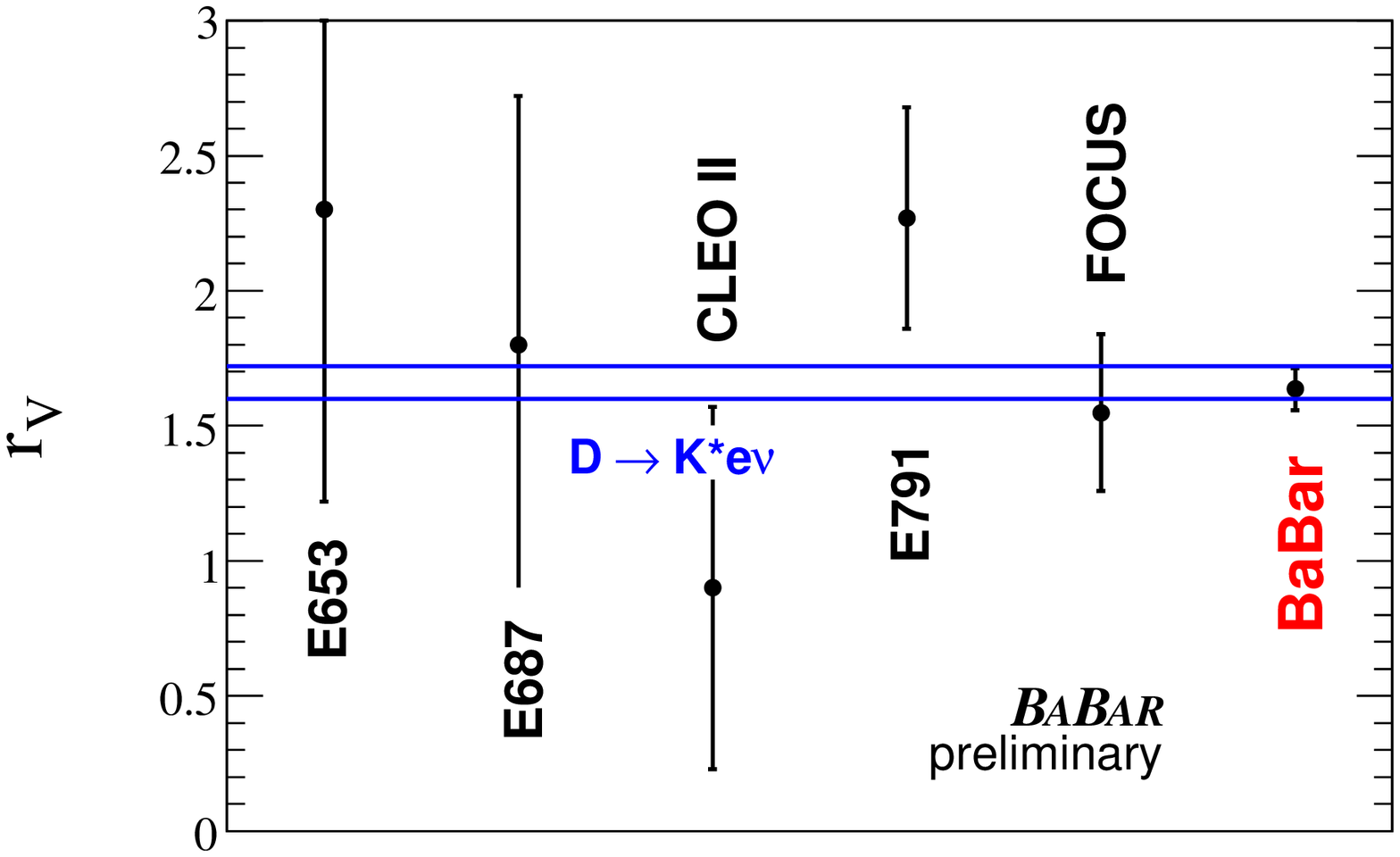,
    width=0.6\textwidth}
  }
  \end{center}
  \caption[]{Results from previous experiments and present measurement
of $r_2$ and $r_V$ in $\Ds \rightarrow \phi e^+ \nu_e $ decays. The error bars represent the statistical and systematic uncertainties added in quadrature.
These measurements for $D_s \to \phi e^+ \nu_e $ decays are compared with the average of similar measurements obtained for  $D \rightarrow K^* e^+ \nu_e $ decays. The $\pm$ one sigma range is indicated by the two parallel lines. 
}
   \label{fig:r2rv}
\end{figure}

The measurements of the parameters $r_V$ and $r_2$ for the semileptonic
decay $\Ds \rightarrow \phi e^+ \nu_e$ now have an accuracy similar
to the one obtained for $D \rightarrow K^* e^+ \nu_e $ decays \cite{ref:wiss}.
This allows meaningful comparisons for the first time, see Table
\ref{tab:comparison} and Figure \ref{fig:r2rv}.
Measurements of $r_V$ for the two decays are in full agreement within the experimental uncertainties.
This was expected theoretically, given the parameterization of Equation \ref{eq:svetla}. However, the measured value is lower than the expectation of 1.8.  Values for $r_2$ differ by 1.5 $\sigma$ between the two decay modes.

\begin{table}[htbp]
\begin{center}
 \caption[]{Comparison between $D/\Ds \rightarrow V e^+ \nu_e$
decays.
  \label{tab:comparison}}
  \begin{tabular}{|c|c|c|}
    \hline
Parameter  &$\Ds \rightarrow \phi e^+ \nu_e$  & $D \rightarrow K^* e^+ \nu_e $ \\
     & (this analysis) & (average value at FPCP06) \\
\hline
 $r_V$ & $1.636\pm0.067\pm0.038$ & $1.66\pm0.06$ \\
    \hline
 $r_2$ & $0.705\pm0.056\pm0.029$ & $0.827\pm0.055$ \\
\hline
  \end{tabular}
 
\end{center}
\end{table}

\section{ACKNOWLEDGMENTS}
\label{sec:Acknowledgments}

We are grateful for the 
extraordinary contributions of our \pep2\ colleagues in
achieving the excellent luminosity and machine conditions
that have made this work possible.
The success of this project also relies critically on the 
expertise and dedication of the computing organizations that 
support \babar.
The collaborating institutions wish to thank 
SLAC for its support and the kind hospitality extended to them. 
This work is supported by the
US Department of Energy
and National Science Foundation, the
Natural Sciences and Engineering Research Council (Canada),
Institute of High Energy Physics (China), the
Commissariat \`a l'Energie Atomique and
Institut National de Physique Nucl\'eaire et de Physique des Particules
(France), the
Bundesministerium f\"ur Bildung und Forschung and
Deutsche Forschungsgemeinschaft
(Germany), the
Istituto Nazionale di Fisica Nucleare (Italy),
the Foundation for Fundamental Research on Matter (The Netherlands),
the Research Council of Norway, the
Ministry of Science and Technology of the Russian Federation, 
Ministerio de Educaci\'on y Ciencia (Spain), and the
Particle Physics and Astronomy Research Council (United Kingdom). 
Individuals have received support from 
the Marie-Curie IEF program (European Union) and
the A. P. Sloan Foundation.


\begin{thebibliography}{99}



\bibitem{ref:KS}
   G.Kopp, G. Kramer, G.A.~Schuler and W.F. Palmer, Z.~Phys.~C~{\bf 48},~327~(1990).

\bibitem{ref:ks2}
J.G. Koerner and G.A. Schuler, Z.~Phys.~C~{\bf 38},~511~(1988);
Erratum-ibid~C~{\bf 41},~690~(1989).\\
M. Bauer and M. Wirbel, Z.~Phys.~C~{\bf 42},~671~(1989).

\bibitem{ref:bad1401}
B. Aubert {\it et al.}, \babar\ Collaboration, \babar\-CONF-06/006, SLAC-PUB-0000, hep-ex/0000, 
submitted to the 33$^{\rm rd}$ International Conference
on High-Energy Physics, ICHEP 06, 26 July - 2 August 2006, Moscow, Russia, and
references therein.

\bibitem{ref:bk}
D. Becirevic and A.B. Kaidalov, Phys.~Lett.~B~{\bf 478},~417~(2000) [arXiv:hep-ph/9904490].

\bibitem{ref:svetlana}
S.Fajfer and J. Kamenik, Phys.~Rev.~D~{\bf 72},~034029~(2005) [arXiv:hep-ph/0506051]
and [arXiv:hep-ph/0601028].


\bibitem{ref:babar}
B.\ Aubert {\em et al.}, \babar\ Collaboration
Nucl.\ Instrum.\ Methods~A~{\bf 479},~1~(2002).

\bibitem{ref:r2}
G. C. Fox and S. Wolfram, Phys.~Rev.~Lett.~{\bf 41},~1581~(1978). 




\bibitem{ref:e653phi}
 K. Kodama {\it et al.}, E653 Collaboration, Phys.~Lett.~B~{\bf 309},~483~(1993).

\bibitem{ref:e687phi}
P.~L.~Frabetti {\it et al.}, E687 Collaboration, Phys.~Lett.~B~{\bf 328},~187~(1994).

\bibitem{ref:e791phi}
E.M. Aitala {\it et al.}, E791 Collaboration, Phys.~Lett.~B~{\bf 450},~294~(1999) [arXiv:hep-ex/9812013].

\bibitem{ref:focusphi}
J.M. Link {\it et al.}, FOCUS Collaboration, Phys.~Lett.~B~{\bf 586},~183~(2004) [arXiv:hep-ex/0401001].

\bibitem{ref:cleophi}
P. Avery {\it et al.} , CLEO Collaboration, Phys.~Lett.~B~{\bf 337},~405~(1994). 

\bibitem{ref:wiss}
J. Wiss, ``Recent results on fully leptonic and semileptonic charm decays'',
FPCP Conference, Vancouver 2006 [arXiv:hep-ex/0605030].



\end{thebibliography}
\end{document}